\documentstyle[12pt]{article}

\font\mybb=msbm10 at 11pt
\def\bb#1{\hbox{\mybb#1}}
\def\ZZ{\bb{Z}}
\def\RR{\bb{R}}

\begin{document}
\thispagestyle{empty}
{\baselineskip=12pt
\hfill CALT-68-1984

\hfill hep-th/9503127

\hfill March 1995

\vspace{1.0cm}}
\centerline{\large \bf String Theory
Symmetries\footnote{Work supported in part by the U.S. Dept. of Energy
under Grant No. DE-FG03-92-ER40701.}}
\bigskip
\bigskip
\centerline{\large John H. Schwarz\footnote{Email: jhs@theory.caltech.edu}}
\medskip
\centerline{\it California Institute of Technology, Pasadena, CA 91125, USA}
\bigskip
\bigskip
\bigskip
\centerline{{\it Presented at the Workshop ``Physics at the Planck Scale''}}

\centerline{{\it Puri, India \quad December 1994}}
\vskip 1.0 truein
\parindent=1 cm

\begin{abstract}
A brief review of the status of duality symmetries in string theory is
presented. The evidence is accumulating rapidly that
an enormous group of duality symmetries,
including perturbative T dualities and non-perturbative S-dualities,
underlies string theory. It is my hope that an understanding of these
symmetries will suggest the right way to formulate non-perturbative
string theory. Whether or not this hope is realized, it has already been
demonstrated that this line of inquiry leads to powerful new tools
for understanding gauge theories and new evidence for the uniqueness of
string theory, as well as deep mathematical results.
\end{abstract}
\vfil\eject
\setcounter{page}{1}
\section{Introduction}
\vglue0.4cm
Duality symmetries in string theory have gained more and more attention in
recent years.  As our understanding increases, it is becoming clear that they
are trying to tell us something important about the structure of string theory.
At first, the focus was on $T$ duality ${}^{1}$ -- or
target-space duality -- which
holds order-by-order in string perturbation theory, though it is
nonperturbative on the string world sheet.  Now the focus is shifting towards
$S$ duality ${}^{2}$ -- or coupling-constant duality --
which is truly nonperturbative.  At the same time, evidence is coming from
several different directions that $S$ and $T$ duality are part of a larger
unified symmetry structure, for which the name $U$ duality has been
coined.${}^{3}$

The history of these ideas goes back to studies of supergravity theories in the
mid 1970s on the one hand,${}^{4,5}$
and to conjectures concerning electric-magnetic
dualities of supersymmetric gauge theories on the other.${}^{6}$
Both of these themes
have been pursued, and carried a great deal further, in the last couple of
years.

Supergravity theories generically contain non-compact global symmetry groups.
The general rule is that the scalar fields of the theory in question
parametrize a symmetric space.  Thus, if the non-compact symmetry group is $G$,
and its maximal compact subgroup is $H$, the scalar fields map the space-time
into the symmetric space $G/H$, and the number of scalar fields is dim $G$
-- dim $H$.  The first supergravity example of this type to be found, $N = 4$
supergravity in four dimensions, is one of the most interesting.  In this case
there are two scalar fields and the symmetric space is $SL(2,\RR)/SO(2)$.
By now, all the possibilities for any number of
supersymmetries and any dimension of space-time have been enumerated.${}^{5}$

It has been clear for some time that instanton effects modify the story in the
quantum theory in a significant way.  For example, in four dimensions certain
of the scalar fields appear as coefficients of $F\tilde{F}$ terms, and,
therefore, the Peccei--Quinn translational symmetry
of the classical theory is broken to (at most) a discrete subgroup in the
quantum theory.  For the specific example of $N = 4, D = 4$ supergravity, this
suggests that the $SL(2,\RR)$ symmetry is broken to an $SL(2,\ZZ)$ subgroup.
Now, as everyone should know, four-dimensional gravity and supergravity
theories are not consistent quantum theories.  But they do represent low-energy
approximations to string theory (for suitable classes of vacua), which probably
is a consistent quantum theory.  These considerations led Font, {\it el al.},
to conjecture that the $SL(2,\ZZ)$ group should be an exact symmetry of
non-perturbative quantum string theory, which they called $S$ duality.${}^{2}$

The conjecture that $S$ duality is an exact symmetry of string theory was a
bold conjecture, because it includes an electric-magnetic duality
transformation as a special case.  Such a transformation sends the electric
charge to the magnetic charge, which is inversely proportional to the electric
charge, because of the Dirac quantization condition.  Thus, it relates weak and
strong coupling, and it is inherently nonperturbative.  This represents an
extension of the Montonen--Olive duality conjecture,${}^{6}$
which relates one theory
with a particular coupling constant to a (possibly different) theory with a
different coupling constant.  In the string theory context, the coupling
constants are given by the expectation values of scalar fields, the duality
transformations are realized as field transformations, and the conjectured
duality can be a symmetry of a single theory.

Once the conjecture of $S$ duality is formulated, we are faced with a
paradoxical situation, which arises from the limitations of our knowledge about
string theory.  What is known, and reasonably well understood, are recipes for
constructing ``classical solutions'' and for adding quantum corrections to any
finite order in perturbation theory.  What is not known, is the equation that
the classical ``solutions'' solve, or a non-perturbative formulation of the
quantum theory.  Given this state of affairs, a natural reaction to the
$S$-duality conjecture would be to declare it ``premature'' on the grounds that
non-perturbative symmetries cannot be tested in a theory that is only known
perturbatively.  Strictly speaking, this is indeed true.  However, if one is
willing to restrict attention to vacua with lots of supersymmetry and make some
mild asumptions about the non-perturbative physics, then some interesting and
non-trivial tests of $S$ duality can be carried out.  All such tests to date
have been successful.  Thus, while $S$ duality has certainly not been proved to
be a string theory symmetry, the situation is quite encouraging.${}^{7,8}$

My point of view is that the purpose of these studies should not be viewed as
verifications of the symmetry, but rather as explanations of how it operates.
They can provide important clues about the structure of non-perturbative string
theory, which could eventually provide guidance in attempts to formulate
non-perturbative string theory.  Perhaps the most important information to be
learned is the fundamental underlying symmetry group of the theory.  Finding
this symmetry is a non-trivial challenge because, for any particular choice of
vacuum (or classical solution), there is a great deal of spontaneous symmetry
breaking.  As we have said, duality symmetries are realized by non-linear
transformations of scalar fields that describe symmetric spaces.  Models of
this type (sometimes called non-linear sigma models) are prototypes for
Goldstone bosons and spontaneous symmetry breaking phenomena.

As has already been indicated, larger pieces of the hidden duality symmetries
of string theory become visible by restricting attention to classical
backgrounds with lots of supersymmetry.  It is also advantageous to consider
low space-time dimensions.  The reason for this is intuitively clear.  The
analysis of duality symmetries is carried out in terms of effective field
theories containing scalar fields associated with symmetric spaces.  However,
in the underlying string theory it is clear that all string modes should have a
democratic status, irrespective of their spins.  By considering ground states
in
low dimension, one limits the possibilities for the spins and forces a large
portion of the spectrum to reside in scalar fields, and therefore more of
the symmetry can be characterized by symmetric spaces.  In the extreme case
where all spatial dimensions have been suitably compactified (or eliminated),
one may have a chance of identifying a very large symmetry group such as
$E_{10}$ or the monster Lie algebra,
or discrete subgroups of these.${}^{9,10}$  This kind
of information could prove crucial in formulating string theory.  The key point
is that the symmetries that are being identified are properties of the
underlying theory, irrespective of the choice of any classical background or
quantum vacuum.  The only purpose of the background or vacuum is to make the
symmetry visible.

Recent work has made it dramatically clear that these studies are valuable even
if they do not serve the purpose that I am advocating.  Studies of
electric-magnetic duality transformations of supersymmetric gauge theories have
clarified the Montonen--Olive conjecture a great deal and lent additional
support to the $S$-duality conjecture in string theory.${}^{11,12}$
Perhaps more
importantly, they have provided remarkable exact non-perturbative results
concerning fundamental issues in gauge theory.  With a slight twist, they have
also had a major impact in the mathematical world.${}^{13}$
These achievements should
convince any sceptic that this is a useful line of inquiry.  It is important to
understand that in these gauge theories the scalar fields of string theory have
been frozen out, {\it i.e.}, replaced by their expectation values.  As a
consequence, the string theory duality symmetry, which was described by a field
transformation, is replaced by a transformation of parameters.  Thus, what one
wishes to demonstrate in these cases is the equivalence of a family of theories
rather than the symmetry of a single theory.  This viewpoint is clearest in the
case of $N=4$ Yang--Mills theories, where the parameters really are fixed
constants (since the theory is finite).  For renormalizable theories in which
parameters run with scale, and dimensional transmutation occurs, there are
additional issues.  These have been addressed in the $N=2$ case in Ref. 12.

\vglue0.6cm
\section{Duality Symmetries in Four Dimensions}
\vglue0.4cm
When the heterotic string theory is toroidally compactified to four dimensions,
in the manner described by Narain,${}^{14}$
two distinct duality groups appear.  One way
of understanding this is to focus attention on the massless scalar modes of the
low-energy effective supergravity theory.  These fields describe the product of
two symmetric spaces.  The $T$ duality symmetric space is $O(6,22)/O(6) \times
O(22)$, and the $S$ duality one is $SL(2,\RR)/SO(2)$.  The 132 scalars that
parametrize the $T$ space have their origin in internal components of the
ten-dimensional metric tensor $g_{\mu\nu}$, antisymmetric tensor $B_{\mu\nu}$,
and 16 $U(1)$ gauge fields $A_\mu^I$ (corresponding to the Cartan subalgebras
of $E_8 \times E_8$ or $SO(32)$).  The two scalars that parametrize  the $S$
space correspond to the dilaton $\phi$ (the ten-dimensional dilaton redefined
by
a term involving the other scalars) and the axion $\chi$ (obtained
by replacing the four-dimensional $B_{\mu\nu}$ by its dual).  The associated
duality symmetries are restricted to the subgroups $G_T = O(6,22;\ZZ)$ and $G_S
= SL(2,\ZZ)$.

The continuous symmetries $O(6,22)$ and $SL(2,\RR)$ are global symmetries of
the classical effective field theory, but the duality subgroups are gauge
symmetries of the string theory.  Thus, the underlying gauge symmetry of string
theory, whether continuous or discrete, should contain these groups as
subgroups.  The meaning of gauge symmetry in string theory ultimately must be
that configurations related by symmetry
transformations are identified as physically identical, and should be counted
only once in the path integral that defines the theory.  Since space-time must
ultimately be a derived concept in string theory, not built into the basic
(background-independent) formulation, it would be meaningless to speak about
space-time dependent transformations, which is the way we are accustomed to
describing gauge symmetries in ordinary gauge theories.

The complete massless spectrum of the toroidally compactified classical
heterotic string contains the scalars discussed above, as well as various other
fields, which will now be enumerated.  The 132 $T$ moduli are conveniently
described by a $28 \times 28$ matrix $M$, belonging to the group $O(6,22)$.
This means that if $L$ is the metric matrix of $O(6,22)$,
$$
M^T LM = L . \eqno (1)$$
The restriction to the coset space $O(6,22)/O(6) \times O (22)$ is achieved by
requiring in addition that
$$
M^T = M . \eqno (2)$$
This
procedure for describing a non-compact symmetric space is completely general.
If (for some other problem) the matrix $M$ is complex, then the restriction to
the coset is given by $M^\dagger = M$.
One can introduce an analogous
$2 \times 2$ matrix ${\cal M}$ belonging to
$SL(2,\RR)$, also taken to be symmetric, to describe the $S$ moduli.
Alternatively, since there are only two $S$ moduli, and the coset space is
K\"ahler, one can describe it by a single complex scalar field
$$
\lambda = \lambda_1 + i \lambda_2 = {\chi} + ie^{-\phi} , \eqno (3)$$
where ${\chi}$ and $\phi$ are the axion and dilaton.

The rest of the massless spectrum depends on the values of the $T$ moduli.
Since $G_T$ is a gauge symmetry of the string theory, the $T$ moduli space is
actually $O(6,22)/O(6) \times O(22) \times G_T$.  The division by $G_T$
introduces orbifold points, which correspond to points of enhanced gauge
symmetry where the massless spectrum is enlarged.  We choose to avoid these
points, restricting $M$ to ``generic points in moduli space.''  Then the
massless spectrum is as follows:  The graviton, described by the canonically
normalized Einstein metric, is invariant under both $S$ and $T$
transformations.  (Note that the ``string metric,'' which differs by a
dilaton-dependent factor transforms under $S$ transformations.)  The
axion-dilaton field $\lambda$ is invariant under $T$ transformations and
undergoes linear fractional transformations
$$
\lambda \rightarrow {a\lambda + b\over c\lambda + d} , \eqno (4)$$
under the $S$ group.  The 132 moduli described by $M$, transform as a
symmetric $28 \times 28$ representation of the $T$ group and
are invariant under the $S$ group.
In addition, there are 28 abelian gauge fields $A_\mu^a$, giving
a gauge group $[U(1)]^{28}$, which form a 28-dimensional representation of
$O(6,22)$.  Finally, there are all the fermions required by supersymmetry,
which we will omit from further consideration.  (They are not an essential
complication.)

We can now write down a Lagrangian that has manifest $T$ symmetry -- the $S$
structure will require additional discussion:
$$
{1\over\sqrt{-g}} L = R - {1\over 2\lambda_2^2} g^{\mu\nu} \partial_\mu
\bar\lambda \partial_\mu \lambda + {1\over 8} g^{\mu\nu} tr (M^{-1}
\partial_\mu M M^{-1} \partial_\mu M) + {1\over 4} L_{ab} F_{\mu\nu}^a
\tilde{G}^{\mu\nu b} , \eqno (5)$$
where
$$
G_{\mu\nu}^a = \lambda_1 F_{\mu\nu}^a + \lambda_2 (ML)_{~b}^a
\tilde{F}_{\mu\nu}^b , \eqno (6)$$
and tilde represents the covariant dual.  The first three terms in the
lagrangian are invariant under the $S$ group $SL(2,\RR)$, but the last one is
not.  However, the equations of motion do have $SL(2,\RR)$ symmetry, so that it
is a classical symmetry.  To see this one, one must require that $G_{\mu\nu}^a$
and $F_{\mu\nu}^a$ from an $SL(2,\RR)$ doublet, and that
$$
\left(\begin{array}{c}
G\\ F
\end{array} \right) \rightarrow \left(\begin{array}{cc} a & b\\ c & d
\end{array} \right) \left(\begin{array}{c} G\\ F \end{array} \right) .
 \eqno (7)$$
Note that this means that $F$ and $\tilde{F}$ get mixed (in a way that depends
on all the moduli), which is the hallmark of an electric-magnetic duality.

The structure of the theory described above can be described in
greater generality, without choosing specific groups.  The discussion that
follows is based on the paper of Hull and Townsend,${}^{3}$
who built on earlier work by Gaillard and Zumino.${}^{15}$
The lagrangian above has a number of scalars $\phi^i$
and the structure
$$
{1\over \sqrt{-g}} L = R - {1\over 2} g_{ij} (\phi) \partial_\mu \phi^i
\partial^\mu \phi^j + {1\over 4} F_{\mu\nu}^I \tilde{G}_I^{\mu\nu} , \eqno
(8)$$
where
$$
G_{\mu\nu I} = a_{IJ} (\phi) F_{\mu\nu}^J + m_{IJ} (\phi) \tilde{F}_{\mu\nu}^J
. \eqno (9)$$
There are $n$ equations of motion and $n$ Bianchi identities for the gauge
fields.  These $2n$ equations can be succinctly written as $d {\cal F} = 0$,
where ${\cal F}$ is the 2$n$-component vector $({\cal F}^I, G_I)$.

Gauss's law can be used to define  $2n$ conserved charges
$$
Z = \oint_\Sigma {\cal F} = (p^I, q_I) . \eqno (10)$$
Here $\Sigma$ is a large two-sphere, and all fields are assumed to approach
constant valus at spatial infinity.  $p^I$ and $q_I$ encode convenient linear
combinations of the $2n$ conserved electric and magnetic charges.  In terms of
these combinations the most general version of the Dirac quantization
conditions for a pair of dyons with charges $(p^I, q_I)$ and $(p^{I\prime},
q'_I)$ in the presence of vacuum $\theta$ angles takes the form
$$
p^I q'_I - p^{I\prime} q_I \in \ZZ . \eqno (11)$$
The significant fact about this formula is its symplectic structure, {\it
i.e.}, it is invariant under $Sp (2n)$ transformations.  This implies that the
allowed charges $(q_I, p^I)$ form a self-dual lattice that is invariant under
$Sp (2n; \ZZ)$.  Without the restriction to integers, the quantization
condition would be violated.  Therefore, the most general global symmetry group
that can arise in a theory of this type must be a subgroup of $Sp (2n; \ZZ)$.
In the heterotic string example we had $n = 28$, and it is true that
$$
SL(2,\RR) \times O(6,22) \subset Sp (56) . \eqno (12)$$
In fact, the fundamental
 {56} of $Sp (56)$ corresponds to $({  2},
{  28})$ for the
subgroup.  Moreover, as required by the Dirac quantization condition,
$$
SL (2, \ZZ) \times O (6,22; \ZZ) \subset Sp (56, \ZZ). \eqno (13)$$
\section{Other Examples}
\vglue0.4cm
Let us start by describing one more example in four dimensions, and then
discuss what happens in other dimensions.  The type II superstring has twice as
much  supersymmetry as the heterotic string, and so when toroidally
compactified to four dimensions, the low-energy effective field theory is $N=8$
supergravity.  The classical global symmetry of this
theory is $E_{7,7}$, which is the maximally non-compact form of $E_7$.  The
fundamental representation of $E_7$ is $56$-dimensional, and the $E_{7,7}$
matrices in this representation are symplectic. Thus, $E_{7,7} \subseteq Sp
(56)$, as required.  It may not be obvious at first how to define a discrete
subgroup $E_7 (\ZZ)$, but Hull and Townsend point out that the appropriate
definition is clearly
$$
E_7 (\ZZ) \equiv E_{7,7} \cap Sp (56, \ZZ) . \eqno (14)$$
This is the four-dimensional duality of the type II string in four
dimensions, and the expected continuous $S$ and $T$ duality groups
are $SL(2,R)$ and $O(6,6)$.  In fact, we are finding more, since
$$
SL(2,R) \times O(6,6) \subset E_{7,7} \eqno (15)$$
is a proper subgroup.  So we have $S \times T\subset U$, where the $U$ duality
group is $E_7 (\ZZ)$.

It is a curious coincidence that both the heterotic and type II theories in
four dimensions have gauge group $[U(1)]^{28}$.  However, there are also some
interesting differences.  In the heterotic case, there are 6 compactified
right-movers and 22 compactified left-movers, so that all 28 electric charges
arise in the elementary string spectrum as Kaluza--Klein and winding-mode
excitations, {\it i.e.}, as discrete internal left- and right-moving momenta.
In the type II case, on the other hand, there are just 6 right-movers and 6
left-movers, so that only 12 of the 28 types of electric charges arise in the
spectrum of elementary string excitations.  Group theoretically, the
${  56}$ of $E_{7,7}$ decomposes under $S \times T$ as
$({  2}, {  12}) + ({  1},
{  32})$, where ${  32}$ is a
spinor representation of $O(6,6)$.  The
$({  2}, {  12})$ describes the 12 electric
charges which are excited in the elementary string spectrum as well as the
corresponding magnetic charges, while
$({  1},{  32})$ describes the remaining electric
and magnetic charges.  Thus, just as for the magnetic charges, these latter
electric charges only appear in the soliton spectrum.

Let us now briefly discuss what happens in other dimensions $(d)$, always
assuming that we are considering toroidally compactified strings.  (Other
geometries will generally
give smaller groups and be more difficult to analyze.)  First
of all, the $T$ duality groups, whose origin is directly tied to the
compactified dimensions, is known to be $O(10-d, 26-d; \ZZ)$ in the heterotic
case and $O(10-d, 10-d; \ZZ)$ in the type II case.  $S$ duality on the other
hand depended on the fact that
a duality transformation applied to $B_{\mu\nu}$ gives rise
to a scalar axion, so it has no counterpart in dimensions greater than four.
However, this is often not the whole story.  The complete $U$ duality group is
larger than $S\times T$ for type II strings in any dimension and for heterotic
strings in dimensions less than four.

In the case of the type II theory, the relevant symmetry is determined by the
corresponding maximal supergravity theory.  The global symmetry of these
theories in $d$ dimensions turns out to be $E_{11-d, 11-d}$.  This has been
demonstrated in detail for the cases $d = 3,4,5$ corresponding to usual Cartan
$E$ groups.  For other values of $d$, the meaning of $E_{11-d}$ is given by
adding or removing dots from the Dynkin diagram in the standard way.  Thus, for
example, $E_{5,5} = SL (5, \RR)$ and $E_{1,1} = SL(2,\RR)$.  The latter is a
symmetry only of the type IIB theory in ten dimensions.  What is more
interesting to contemplate is the larger symmetries that this formula suggests
for $d < 3$.  $E_9$ is the affine extension of $E_8$ and $E_{10}$ is a poorly
understood hyperbolic Lie algebra.  Nicolai has presented some evidence that
$E_9$ is a global symmetry of the classical two-dimensional supergravity
theory,${}^{16}$
though work that I am currently doing suggests that this may not be
precisely correct.${}^{17}$
In any case, what is of greater importance is to identify
the discrete subgroup that should describe a symmetry of the quantum theory.
Understanding $E_{10}$ and figuring out whether it has something to do with the
symmetry of string theory is an even greater challenge.

In the case of the heterotic string theory, dimensions less than four are also
interesting.  It was shown a long time that the three-dimensional supergravity
theory has an $O(8,24)$ symmetry.${}^{18}$
This is an extension of the expected $S \times T$ symmetry:
$$
O(8,24) \supset   SL(2,\RR) \times O (7,23). \eqno (16)$$
Recently, Sen has explained that the string theory duality symmetry in this
case
is $O(8,24; \ZZ)$,${}^{19}$
which can be viewed as a $U$ duality extension of $SL(2,\ZZ)
\times O (7,23;\ZZ)$.  The next step is to investigate what
happens on reduction to two dimensions.${}^{20,21}$
The general rule for symmetric space
models, at least roughly, is believed to be that if the symmetry group in three
dimensions is $G$, reduction to two dimensions will give a classical theory
whose symmetry is $\hat G$, the affine extension of $G$.  I find
that actually the classical symmetry is a large subgroup
of $\hat G$, which I call $\hat G_H$, where $H$ is the maximal compact subgroup
of $G$.${}^{17}$  Sen has recently reported impressive progress in
identifying the discrete duality group of the toroidally compactified
heterotic string in two dimensions.${}^{22}$
Once a satisfactory understanding of the situation in two dimensions is
achieved, we may build up the courage to investigate the hyperbolic algebra and
its discrete subgroup that are expected to appear upon reduction to one
dimension.
\vglue0.6cm
\section{Solitons and String Excitations}
\vglue0.4cm
Let us now return to the heterotic string theory toroidally compactified to
four dimensions, for which the duality group is $SL(2;\ZZ) \times O(6,22;\ZZ)$
and consider the spectrum of charged excitations.  The great advantage of
having $N=4$ supersymmetry is that the supersymmetry algebra contains central
charges whose origin can be understood in terms of internal components of the
ten-dimensional momenta.  Thus, since supersymmetry is associated with
right-movers, there are six complex central charges corresponding to
right-moving electric and magnetic charges.  The consequence of this is that
one can derive a lower bound on the $(mass)^2$ of any charged state, known as
the Bogomol'nyi bound.  In terms of the charges $p^I$ and $q_I$ introduced in
section two, it takes the form
$$
M^2 \geq {1\over 2} (p^I, q_I){\cal M}^{(0)} (M^{(0)} + L)_{IJ} \left(
\begin{array}{c} p^J \\ q_J \end{array} \right), \eqno (17)$$
where an overall constant factor can be absorbed in the definition of the
string scale $\alpha'$.  ${\cal M}^{(0)}$ is a $2\times 2$ matrix representing
the asymptotic values of the $S$ moduli and $M_{IJ}^{(0)}$ is a $28\times 28$
matrix representing the asymptotic values of the $T$ moduli.  The combination
$M^{(0)} + L$ projects out those charge contributions that come from
right-movers.

This inequality must hold so long as the $N=4$ supersymmetry remains unbroken.
What is more interesting is that when it is realized as an equality the algebra
has ``short representations,'' which are not possible otherwise. The basic
``massive'' representative of $N=4$ supersymmetry is
256-dimensional, whereas the basic ``massless'' one is 16-dimensional.  The
latter is very familiar from $N=4$ super Yang--Mills theory and from the ground
state structure of superstrings (8 NS states and 8 R states).  So,
if we identify a set of states that in some approximation saturate
the Bogomol'nyi bound, then they must continue to do so in the exact quantum
theory.  The only assumptions that enter this argument are that $N=4$
supersymmetry remains an exact symmetry and that the full theory does not have
different phases (e.g., one that confines),
which could invalidate the argument.
These mild assumptions allow us to draw exact non-perturbative conclusions
about a theory we don't even know!  (G. `t Hooft found this remark to be very
provocative.)

Let us now examine the spectrum of elementary string excitations.  These are
entirely electric, of course.  As usual, because of the phenomenon of
level-matching, there are two formulas for the mass of each state, one in terms
of left-movers and one in terms of right-movers.  These are
$$
M^2 = {1\over\lambda_2^{(0)}} \left({1\over 2} p_L^2 + N_L - 1\right)
= {1\over\lambda_2^{(0)}} \left({1\over 2} p_R^2 + N_R - \delta\right), \eqno
(18)$$
where $\lambda_2^{(0)}$ is the asymptotic value of $\lambda_2 = e^{-\phi}$, and
the internal momentum contributions are
$$
p_L^2 = {1\over 2} p^I (M^{(0)} - L)_{IJ} p^J \eqno (19)$$
$$
p_R^2 = {1\over 2} p^I (M^{(0)} + L)_{IJ} p^J . \eqno (20)$$
$N_L$ and $N_R$ are the usual oscillator-excitation eigenvalues, and $\delta$
is the superstring zero-point energy (${1\over 2}$ in the NS sector and $0$ in
the R sector).  The factor $\lambda_2^{(0)}$, which is essential (and possibly
unfamiliar), appears because masses are being measured with respect to the
Einstein metric rather than the string metric, as is often assumed.  Now, the
way to saturate the Bogmol'nyi bound is clear.  One must take $N_R = 0$, since
only by taking the ground state of the right-moving oscillators can one obtain
a ``short'' 16-dimensional representation of the supersymmetry algebra.  Of
course, this gets tensored with whatever left-moving contributions occur.

Setting $N_R = 0$ and using the level-matching condition, we see that the most
general elementary string excitations saturating the Bogomol'nyi bound are
characterized by the equation
$$
N_L = 1 + {1\over 2} (p_R^2 - p_L^2) = 1 + {1\over 2} p^I L_{IJ} p^J. \eqno
(21)$$
Since the charge vectors $p^I$ belong to an even self-dual lattice, with metric
$L_{IJ}$, this is guaranteed to be an integer.  There are string states
corresponding to each distinct way of solving this equation (tensored with the
16-dimensional supermultiplet of right-movers).  $S$ duality, if it really is a
symmetry, requires the existence of a rich spectrum of magnetically charged
partners for these particles.  Since none of these are in the spectrum of
elementary excitations, they must all arise as solitons.

The simplest case is when there are no excitations of left-moving oscillators.
In this case $p^2 = p^I L_{IJ} p^J = - 2$, and the complete charge vector is
$(p^I, 0)$.  However, $S$ duality transforms this to the
vector $(a p^I, cp^I)$, where $a$ and $c$ are relatively prime integers.  Since
these will also belong to a short supermultiplet, they must also saturate the
Bogomol'nyi bound and, therefore, eq. (17) tells us what their masses must
be.  The existence of these states has been
discussed in detail by Sen.${}^{7}$  He argues that the states with $N_L = 0$
and $c = 1$ can be identified with BPS
monopoles and dyons (appropriately generalized to the present context), and
that they occur with precisely the required multiplicities and other
properties.  The required states with $c > 1$ are much more difficult to
analyze.  Sen has argued that the $S$-duality conjecture requires that the
moduli space of $c$ BPS monopoles should have one, and only one, normalizable
harmonic form.  Such a form is necessarily self-dual (or anti-self-dual).  He
then proceeded to prove that this is the case for $c = 2$.${}^{23}$
I am informed that
the case $c > 2$ is being investigated by Segal, and that he is making good
progress towards establishing the desired result.

The preceding is mathematically very challenging, even though it only concerns
$N_L = 0$ states.  The next level, $N_L = 1, p^2 = 0$ states include
excitations with spins ranging up to $J = 2$, and certainly probe aspects of
the theory that go well beyond ordinary field theory.  Some of these states,
which are identified with $H$ monopoles,${}^{24}$
have been investigated by Gauntlett and Harvey.${}^{25}$
They have made impressive progress, but their work still
appears to be somewhat inconclusive.
\vglue0.6cm
\section{Conclusion}
\vglue0.4cm
The study of string duality symmetries is proving to be a fascinating subject
that is opening up many new avenues of research.  These have
far-reaching implications for four-dimensional gauge theories and fundamental
mathematics, as well as for the study of string theory itself.  The latter is
certainly the most challenging problem, however.  The hope is that when the
complete group of duality symmetries is identified and understood it will
provide the key to obtaining a complete formulation of string theory.  Whether
or not this goal will ever be achieved remains to be seen, but it has already
been demonstrated that we will learn a great deal in the attempt.
\vglue0.6cm

\end{document}